\documentclass[aps,pra,twocolumn]{revtex4-1}
\usepackage{amsmath}
\usepackage{epsfig}
\usepackage{graphicx}
\usepackage{color}
\usepackage{amssymb}
\usepackage{epstopdf}
\usepackage{diagbox} 
\usepackage[utf8]{inputenc}
\usepackage[T1]{fontenc}
\usepackage[colorlinks=True, linkcolor=blue, citecolor=blue]{hyperref}
\begin{document}
\title{Superfluidity fraction of few bosons in an annular geometry in the presence of a rotating 
weak link}
\author{Alex V. Andriati$^{1}$}\thanks{andriati@if.usp.br}
\author{Arnaldo Gammal$^1$}\thanks{gammal@if.usp.br}
\affiliation{$^{1}$Instituto de F\'{i}sica, Universidade de S\~{a}o Paulo, 05508-090 S\~{a}o Paulo, 
Brazil.}
\date{\today}
\begin{abstract}
We report a beyond mean-field calculation of mass current and superfluidity fraction for a system 
of few bosons confined in a ring geometry in the presence of a rotating weak link induced by a potential 
barrier. We apply the Multiconfiguration Hartree Method for bosons to compute the ground state of 
the system and show the average superfluidity fraction for a wide range of interaction strength and 
barrier height, highlighting the behavior of density correlation functions. The decrease of superfluidity 
fraction due to the increase of barrier height is found whereas the condensation fraction depends 
exclusively on the interaction strength, showing the independence of both phenomena.
\end{abstract}

\flushbottom

\maketitle
\section{Introduction}

The concepts of superfluidity \cite{Khalatnikov,Leggett1998,RevModPhys.Legget1999} and Bose-Einstein 
condensation \cite{Cornell,Pethick} have dominated the research of cold bosonic systems. The presence of one do not necessarily imply the other, whereas superfluidity 
is related to dissipationless flow due to a minimum required energy to create excitations, 
Bose-Einstein condensation is characterized by a single macroscopically occupied state. Superfluid 
systems with very small condensation fraction around $10\%$, like liquid Helium, are widely 
known \cite{Penrose,Sears}, thereby characterizing independent effects.

Nevertheless, many reports explore the superfluidity features of a Bose-Einstein condensate (BEC), 
as dilute cold bosonic gases are able to present both phenomena simultaneously \cite{Ketterle1999}. 
Specially, persistent flow, hallmark of superfluidity, has been  reported for a BEC 
trapped in a ring shape format early in \cite{PhysRevLett.99.260401} and later in 
\cite{TunableWeakLink} in the presence of a tunable weak link. This boosted the interest to 
quantitatively study all the properties for the system due to a possible connection and quantum 
analogy with superconducting quantum interference devices (SQUID) \cite{SQUIDhandbook}, that was 
experimentally implemented \cite{PhysRevLett.111.205301} generating Josephson 
junctions \cite{PhysRevLett.95.010402}.

Many works studied several properties on imposing rotation for a BEC confined in a ring shape 
geometry, observing hysteresis (``swallow tail 
loops'') \cite{Mueller2002,CampbellNature2014,PhysRevA.87.013619,Syafwan2016}, excitation 
mechanisms \cite{PhysRevA.88.063633,PhysRevA.99.043613,PhysRevA.95.021602}, spin 
superflow \cite{PhysRevLett.119.185302,PhysRevLett.110.025301} and superfluid 
fraction \cite{Hadzibabic2010}. Although the theoretical studies rely mostly on the 
Gross-Pitaevskii (GP) equation that set a clear limitation on controlling the 
interactions to suppress the depletion from the condensate \cite{PhysRevLett.119.190404,
PhysRevLett.117.235303}, this has changed in the past few years with the development of novel 
methods able to compute many-body observables and assure correctness for a wider range of 
interaction values.

The employment of new methods pave a way to study independently the condensation phenomena and the 
superfluidity for a system of cold bosonic atoms and to sweep a wider range for interaction 
strength since depletion is included on the description. Moreover, they enable us a deep 
understanding of the physical system through new many-body quantities unseen in the mean field 
formalism, like correlations that has gained importance due to experimental measures in the past 
decade for cold atomic clouds \cite{Navon167,Glauber,10.1038/nphys2632,PhysRevLett.118.240402}.

Specifically, the Multi-Configuration Time Dependent Hartree method for Bosons (MCTDHB) 
\cite{MCTDHBderivation} has dragged attention for its quite straightforward generalization of the GP 
equation, since it is still based on variational principle but with more single-particle states 
(also known as orbitals) the atoms can occupy, therefore allowing the expansion of the many-body 
state in a configuration basis (Fock states) with each possible configuration expressed by a 
well-defined occupation number of the single-particle states. This procedure truncate the Hilbert 
space, whereas the coefficients of the basis expansion and the orbitals are determined by minimizing 
the action, enforcing an optimized basis. The MCTDHB has shown to be a powerful tool for 
many applications like bosons in optical lattices \cite{PhysRevA.97.043625}, quench dynamics 
\cite{PhysRevA.92.033622} and other applications 
\cite{PhysRevX.9.011052,PhysRevA.94.063648,PhysRevA.93.063601} as well as a version for fermions 
\cite{PhysRevA.93.033635}. 

In the present report, we study beyond mean field the superfluidity fraction of a gas of 
bosons at zero temperature in the presence of a tunnable weak link moving in a periodic system (an 
effective ring), for a small number of atoms, using the MCTDHB to exploit strong 
interactions and to show the loss of the superfluidity fraction under a wide range of the physical 
parameters. Explanations for the superfluid properties are studied here through correlation of the 
bosonic field operator in the absence of rotation, which is directly related to the tunneling 
amplitude in the neighborhood of the barrier, therefore allowing a prediction if the atoms would be 
dragged when some rotation is applied.

\section{Model and methods}
\label{secII}

The specific form of a barrier is generally unknown from an experimental perspective, though we 
must be able to define it through its thickness and height. As most of the experiments use lasers 
to physically implement a barrier \cite{TunableWeakLink}, the height in the model plays the 
principal role as it is directly related to laser intensity, and the thickness can be fixed in 
a first moment. Indeed, an approach based on Dirac delta function for the barrier has been
reported \cite{AnnaMinguzzi}, which implies zero thickness. In any case, for a barrier rotating 
with velocity $v$, in the laboratory frame we thus have the one-body term of the Hamiltonian in the 
general form
{\small
    \begin{equation}
        \hat{h}(t) = - \frac{\hbar^2}{2 m} \frac{\partial^2}{\partial \bar{x}^2} +
        U(\bar{x} - vt) \ ; \quad \bar{x} \in (-\pi R, \pi R] ,
        \label{eq:labframe}
    \end{equation}
}

\noindent for a ring of radius $R$. The two-body part is assumed to be described by an effective 
contact interaction $V(\bar{x} - \bar{x}') = g_{1\mathrm{D}} \delta(\bar{x} - \bar{x}')$, 
where $g_{1\mathrm{D}}$ is related to the transverse harmonic trap frequency and the scattering 
length of the atoms. Using a unitary transformation to move to the rotating frame, the time 
dependence of Eq. \eqref{eq:labframe} is removed, resulting in the following many-body 
Hamiltonian in the second quantized formalism
{\small
    \begin{multline}
        \mathcal{H} = \int_{-\pi R}^{\pi R} \!\!\!\!\! \mathrm{d}x \ \Psi^{\dagger}(x) \left[ 
        \frac{\hbar^2}{2 m}
        \left( i \frac{\partial}{\partial x} + \frac{m v}{\hbar} \right)^2 
        + U(x) \right] \Psi(x) \ + \\
        \frac{g_{1\mathrm{D}}}{2} \int \! \! \! \mathrm{d}x \
        \Psi^{\dagger}(x) \Psi^{\dagger}(x) \Psi(x) \Psi(x) ,
        \label{eq:secondquantized}
    \end{multline}
}

\noindent where $x = \bar{x} - vt$.

The MCTDHB is developed assuming a truncated Hilbert space where the many-body state is a 
superposition of all possible configurations $N_c$  of $N$ particles distributed over $M$ 
single-particle states, such that we can write
{\small

    \begin{equation}
        | \Psi(t) \rangle \ \doteq \ \sum_{\alpha = 1}^{N_c} C_{\alpha}(t) | \vec{n}^{(\alpha)}
        \rangle , \quad N_c = \binom{N + M - 1}{ M - 1} ,
        \label{eq:manybody-state}
    \end{equation}
}

\noindent where a valid configuration $| \vec{n}^{(\alpha)} \rangle$ is a Fock state where 
$\sum_{i}^{M} n_i^{(\alpha)} = N$, $\forall \alpha \in \mathbb{N} \ | \ 1 \leq \alpha \leq N_c$. 
Furthermore, the occupation number refers to a set of single-particle states $\{ \phi_k(x,t) \ | \
\int \!\! \mathrm{d}x \phi_l^{*}(x,t) \phi_k(x,t) = \delta_{lk}, \forall k,l = 1, ..., M \}$. Using
this \emph{Ansatz}, the time-dependent equations can be extracted from a minimization of the action with
respect to the coefficients $C_{\alpha}$ in Eq. \eqref{eq:manybody-state} and the single-particle 
state, with the action defined by
{\small
\begin{multline}
    \mathcal{S} \Big[ \mathbf{C} , \{ \phi_k, \phi_k^{*} \} \Big] = 
    \int \!\!\! \mathrm{dt} \bigg[ \langle \Psi(t) | \dot{\Psi}(t) \rangle \ - \ 
    \langle \Psi(t) | \mathcal{H} | \Psi(t) \rangle\ \\
    - \sum_{k,l=1}^{M} \mu_{kl}(t) \langle \phi_k | \phi_l \rangle _t \bigg],
\end{multline}
}

\noindent where the $\mu_{kl}$ are introduced as Lagrangian multipliers to maintain orthonormality 
of the single-particle states. The variational principle conducts to $M$ nonlinear coupled partial 
differential equations for the set $\{ \phi_k(x,t)\}$ and a system of $N_c$ ordinary differential 
equation for the coefficients $C_{\alpha}$ \cite{Hans-Meyer,MCTDHBderivation}. It is worth 
 mentioning that the GP equation is a special case where we have just one possible configuration $| 
\vec{n} \rangle = | N, 0, ..., 0\rangle$, that yield for the macroscopically occupied orbital 
$\phi(x,t)$ the equation
{\small
\begin{equation}
    i \hbar \frac{\partial \phi}{\partial t} = \Big[ \hat{h}' + g (N-1) |\phi(x,t)|^2 \Big]
    \phi (x,t) ,
    \label{eq:gp}
\end{equation}
}

\noindent with $\hat{h}' = \hbar^2 /2 m \left( i \partial / \partial x + m v /\hbar \right)^2 + 
U(x)$ the one-body Hamiltonian in the rotating frame.

For numerical simulation purposes, we assume the following system of units: length measured in units 
of $(\pi R)$, probability/particle density in unis of $(\pi R)^{-1}$ and energy by $\hbar \zeta$ 
where $\zeta = (\hbar / 2 m \pi^2 R^2)$. Moreover, we introduce the dimensionless parameters $\Omega 
= m R v / \hbar$ and $\gamma = 2 m \pi R g_{1\mathrm{D}} / \hbar ^2$. All these transformations 
yield the following orthonormal condition for the set of orbitals: $\int_{-1}^{1} \!\! \mathrm{d} x 
\phi_l^{*}(x,t) \phi_k(x,t) = \delta_{lk}$. Here we developed our own codes to solve the 
MCTDHB equations with periodic boundary conditions. Our codes were extensively tested, matching results of the examples of the code available in Ref.~\cite{MCTDHpackage}, that has produced many results until now \cite{PhysRevA.93.033635, PhysRevA.93.063601, PhysRevA.94.063648, PhysRevX.9.011052, PhysRevA.92.033622}.

\section{Periodicity in energy spectrum and definition of superfluidity fraction}
\label{secIII}

In the absence of a barrier, 
the single-particle energy levels 
as function of $\Omega$ are parabolas 
given by $E_j / (\hbar \zeta) = (j - \Omega)^2 \pi^2$, each one defined by the winding number of the 
phase ($j$), 
centered at $\Omega_j = j$, and crossing each other at 
$\tilde{\Omega}_j = (j + 1/2) ~$\cite{PhysRevA.87.013619}.

\begin{figure}[tbp]
    \centering
    \includegraphics[scale=0.95]{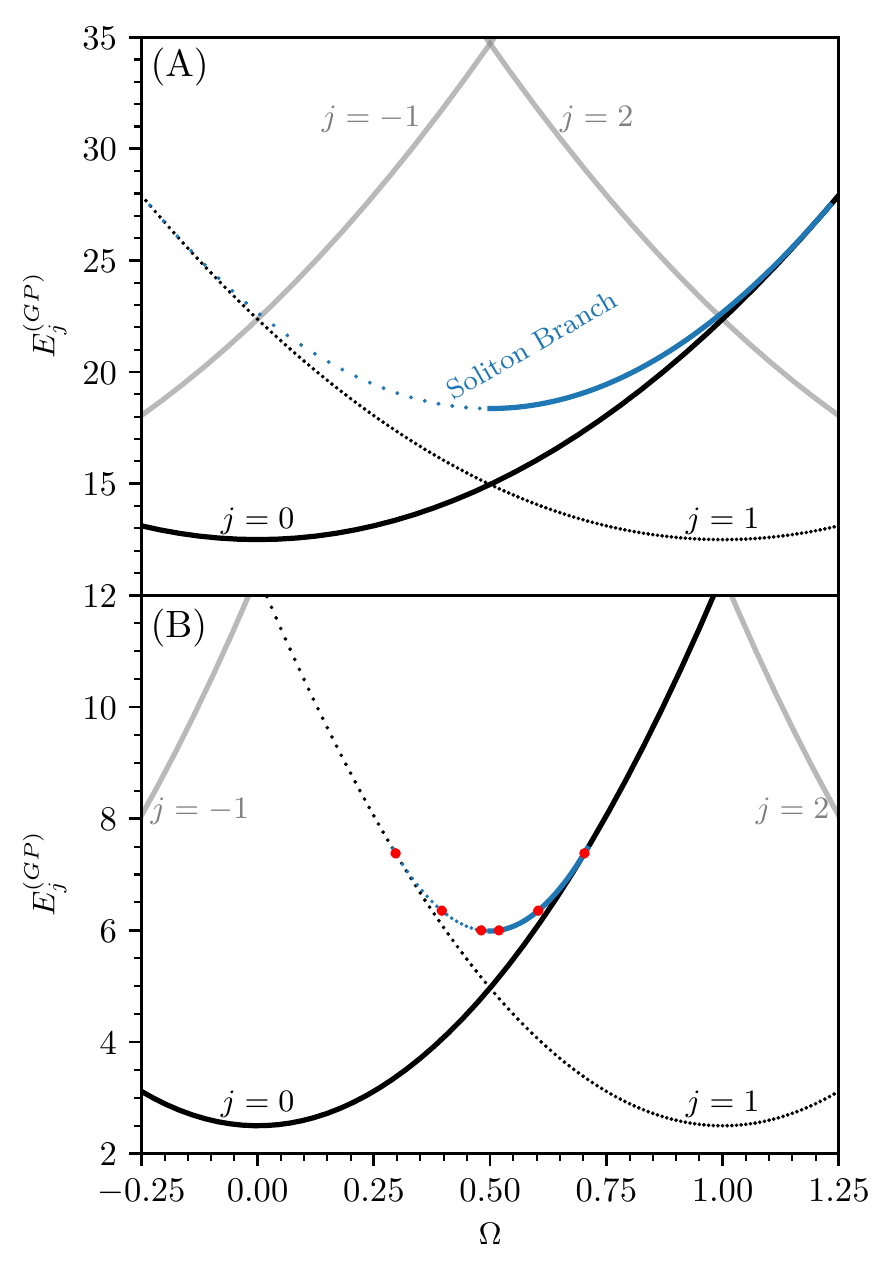}
    \caption{(Color online) Energy per particle from GP equation as a function of $\Omega$ for different winding 
numbers ($j$) with (A) $\gamma (N-1) = 50$ and (B) $\gamma (N-1) = 10$. In both figures the soliton 
solutions energy are depicted in blue, whose connect the parabolas with $j = 0$ to $j = 1$, shown 
by dotted and full line respectively. Other values of winding numbers are shown in gray.}
    \label{fig:nobarrier_energy}
\end{figure}

\begin{figure}[t]
    \centering
    \includegraphics[scale=0.96]{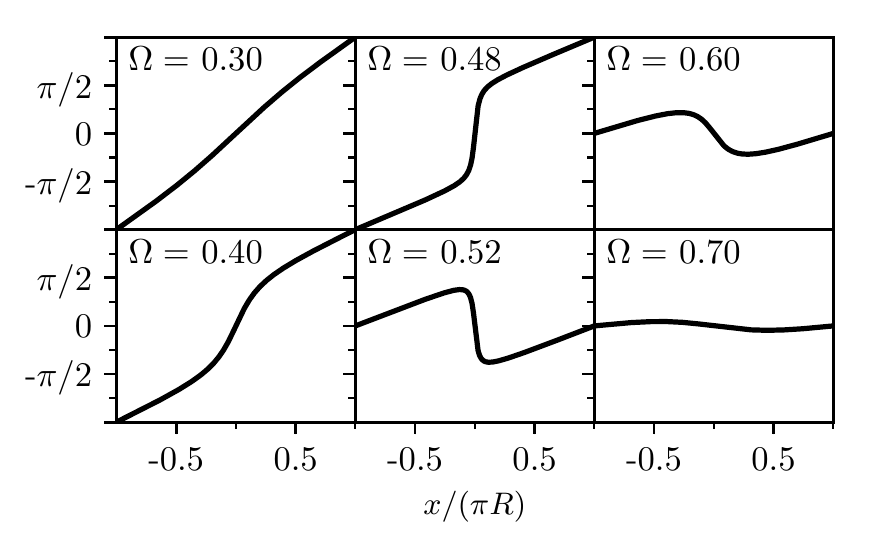}
    \caption{Phase profile of soliton solution for some values of $\Omega$ corresponding to the red 
    dots in \ref{fig:nobarrier_energy}. An abrupt transition occurs at $\Omega = 0.5$ that implies 
    a transition in the winding numbers.}
    \label{fig:nobarrier_soliton_phases}
\end{figure}

As a first approach, we use the GP equation once the interaction is included in the description. 
In 
this case, still in the absence of the barrier, there are two kinds of analytical solutions, one 
with constant density, which results in the same energy of single-particle case, with the 
addition of an interaction contribution, yielding $E^{(GP)}_{j} / (\hbar \zeta) = (j - \Omega)^2 
\pi^2 + \gamma (N-1) / 4$ as average energy per particle. The other is a soliton given in terms of 
Jacobi elliptic functions \cite{PhysRevA.62.063610,Sato2016} that exists for a finite range of 
values of $\Omega$, whereas the extension of this range depends on the interaction strength. 
Fig.~\ref{fig:nobarrier_energy} shows an energy landscape of the analytical solutions of the GP 
equation with the soliton solution energy connecting two parabolas from constant density solutions, 
where the dotted lines have winding number 1 and the filled line 0.

From Fig.~\ref{fig:nobarrier_soliton_phases} we check that the soliton solution is responsible for 
the transition between different winding numbers and a discontinuity occurs in $\Omega = 0.5$ in its 
phase. Moreover, the soliton connects the values of angular momentum \cite{GuilleumasHysteresis} and 
this connection between the two lines is related to an hysteretic behavior by the presence of a 
``swallow-tail'' loop \cite{GuilleumasHysteresis,Mueller2002,CampbellNature2014,PhysRevA.87.013619}. 
The soliton branch in Fig.~\ref{fig:nobarrier_energy} is an excited state \cite{PhysRevA.87.013619}, 
however, it will not be discussed here, since the aim of the present work is to measure the 
superfluidity fraction for the ground state. In addition, Fig.~\ref{fig:nobarrier_energy} reveals 
that the ground state energy has a periodic behavior with respect to the rotation $\Omega$, with 
kinks where the parabolas cross each other at $\tilde{\Omega}_j = (j + 1/2)$. This periodic 
structure remains even in the presence of a barrier as will be shown later. An important fact is 
that we can relate the mass current circulation with the energy, and use this periodicity to 
understand what happens with the current under action of fast rotating barriers.

Here we start a derivation of mass current by looking at the time variation of the number of atoms 
within the range $[x_1 , x_2] \subseteq [-\pi R, \pi R]$, as
{\small
\begin{equation}
    \frac{\mathrm{d}}{\mathrm{d} t}
    \int_{x_1}^{x_2} \!\!\!\!\!\! \mathrm{d} x \ \langle \Phi(t) |
    \hat{\Psi}^{\dagger}(x) \hat{\Psi}(x) | \Phi(t) \rangle = 
    \frac{i}{\hbar} \int_{x_1}^{x_2} \!\!\!\!\!\! \mathrm{d} x \ \langle \big[ \mathcal{H},
    \hat{\Psi}^{\dagger}(x) \hat{\Psi}(x) \big] \rangle_t ,
    \label{eq:TimeVariationOfParticles}
\end{equation}
}

\noindent where $\langle \cdot \rangle_t$ means the expectation value for an 
arbitrary many-body state $| \Phi (t) \rangle$. Using Eq. \eqref{eq:secondquantized} with the 
usual commutation relation for the boson field operator $[\hat{\Psi}(x),\hat{\Psi}^{\dagger}(x')]
= \delta(x - x')$ to evaluate the commutator of the Hamiltonian with the density operator, the only 
terms that contribute are those carrying a derivative, and yields
{\small
\begin{multline}
    \big[ \hat{\Psi}^{\dagger}(x) \hat{\Psi}(x), \mathcal{H} \big] = - \frac{\hbar^2}{2m}
    \left( \hat{\Psi}^{\dagger}(x) \frac{\partial^2 \hat{\Psi}(x)}{ \partial x^2} -
    \frac{\partial^2 \hat{\Psi}^{\dagger}(x)}{ \partial x^2} \hat{\Psi}(x) \right) \\
    + i \hbar v \left( \hat{\Psi}^{\dagger}(x) \frac{\partial \hat{\Psi}(x)}{ \partial x} +    
    \frac{\partial \hat{\Psi}^{\dagger}(x)}{ \partial x} \hat{\Psi}(x) \right) .
    \label{eq:density_H_commutator}
\end{multline}

}

It is straightforward to factor out the derivative with respect to $x$, and further 
using Eq. \eqref{eq:density_H_commutator} in Eq. \eqref{eq:TimeVariationOfParticles} yields
{\small
\begin{equation}
    \frac{\mathrm{d}}{\mathrm{d}t} N([x_1,x_2];t) = -
    \Big[ \langle \hat{J} (x_2) \rangle_t - \langle \hat{J} (x_1) \rangle_t \Big]
\end{equation}
}

\noindent where $ N([x_1,x_2];t) = \int_{x_1}^{x_2} \mathrm{d} x \langle \Phi(t) | 
\hat{\Psi}^{\dagger}(x) \hat{\Psi}(x) | \Phi(t) \rangle$ is introduced and the particle number 
current operator $\hat{J}(x)$ is given by
{\small
\begin{multline}
    \hat{J}(x) = - \frac{i \hbar}{2m}
    \left( \hat{\Psi}^{\dagger}(x) \frac{\partial \hat{\Psi}(x)}{ \partial x} -
    \frac{\partial \hat{\Psi}^{\dagger}(x)}{ \partial x} \hat{\Psi}(x) \right) \\
    - v \hat{\Psi}^{\dagger}(x) \hat{\Psi}(x) .
\end{multline}
}

The reduced single-particle density matrix (1-RDM) defined by $n^{(1)}(x,x';t) \doteq \langle 
\hat{\Psi}^{\dagger}(x') \hat{\Psi}(x) \rangle_t $ \cite{Glauber} has as a set of eigenvalues and 
eingenstates defined by the solution of $\int_{-\pi R}^{\pi R} \! \mathrm{d} x \ n^{(1)}(x,x';t) 
\psi(x',t) = \mathcal{N}(t) \psi(x,t)$, with $\mathcal{N}(t)$ the average occupation number in the 
eigenstate $\psi(x,t)$, here also called as \emph{natural orbital}. Using these natural orbitals to 
express the reduced single-particle density matrix $n^{(1)}(x,x';t)$ allows us to express the current as a 
superposition, $\langle \hat{J}(x) \rangle_t = \sum_{k} j_k(x,t)$, where
{\small
\begin{equation}
    j_k = - \left[ \frac{i \hbar}{2m} \left( \psi_k^{*} \frac{\partial \psi_k}{\partial x} - 
    \psi_k \frac{\partial \psi_k^{*}}{\partial x} \right) + v|\psi_k|^2 \right] \mathcal{N}_k ,
\end{equation}
}

\noindent with the position and time arguments omitted.

For the ground state, the current $\langle \hat{J}(x) \rangle_t$ must be independent of position and 
time, because the density is not time-dependent. If we further average it over a period in the 
counter direction of the barrier velocity, yields
{\small
\begin{equation}
    \langle \rho_s \rangle (v) = \tau \frac{1}{2\pi R} \int_{\pi R}^{-\pi R} \!\!\!\!\!
    \mathrm{d} x \ \left( \frac{\langle \hat{J} \rangle}{N} \right) ,
    \label{eq:currentJ}
\end{equation}
}

\begin{figure}[tbp]
    \centering
    \includegraphics[scale=0.95]{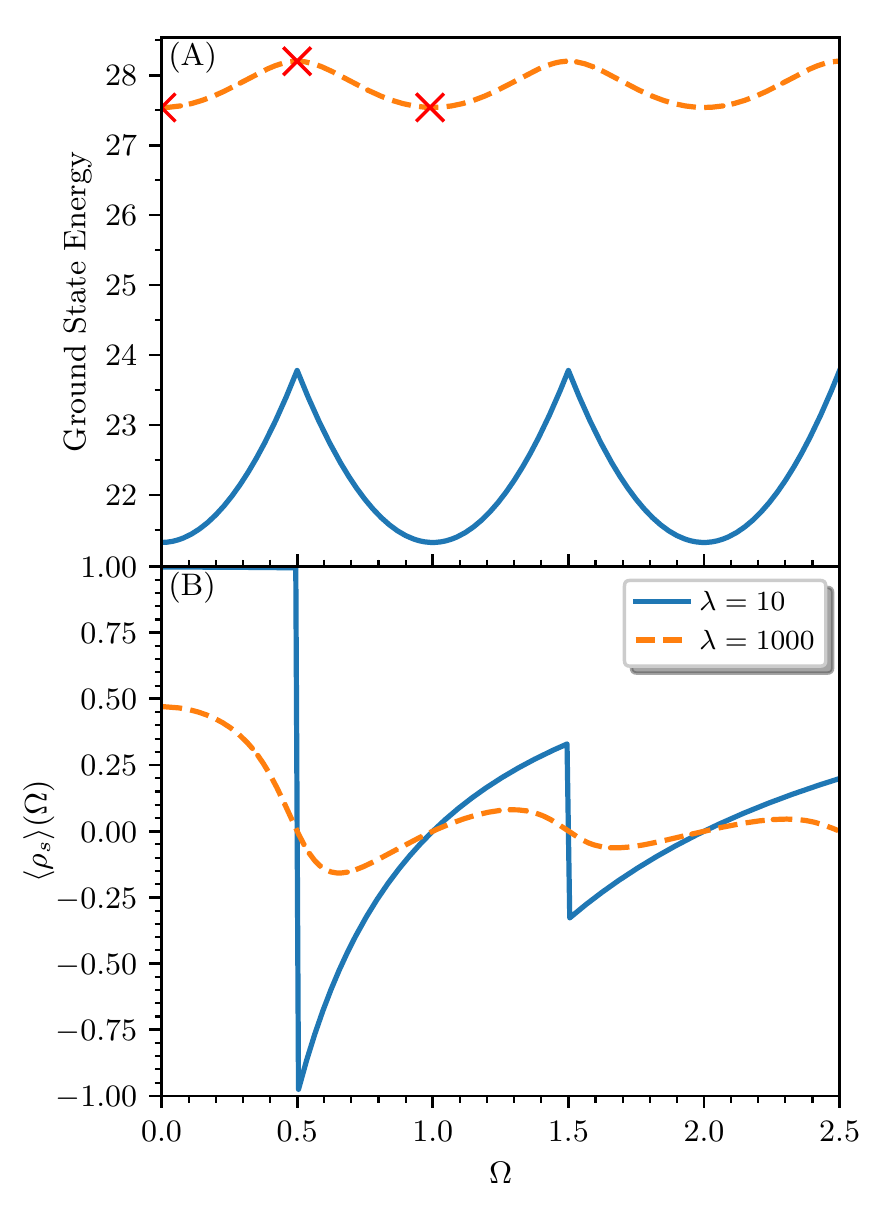}
    \caption{(Color online) Ground state energy and current fraction for 11 particles as a function of 
    dimensionless rotation velocity $\Omega$ in the rotating frame. The ground state energy remains 
    periodic as it was in Fig.~\ref{fig:nobarrier_energy} but with a different landscape 
    depending on the barrier height $\lambda$, and this periodicity implies a decrease on the 
    current fraction for fast rotating barriers. Dimensionless interaction strength parameter 
    used was $\gamma = 10$.}
    \label{fig:CurrentXRotation}
\end{figure}

\begin{figure}[tbp]
    \centering
    \includegraphics[scale=0.96]{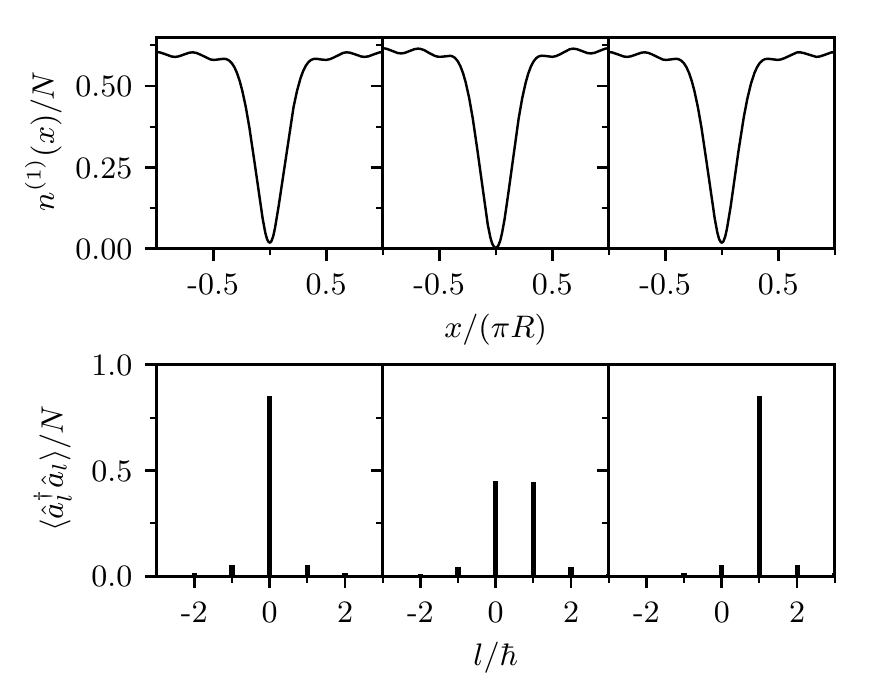}
    \caption{Probability distribution for position(upper panel) where $n^{(1)}(x) \equiv n^{(1)}(x,x) = \langle 
    \hat{\Psi}^{\dagger}(x) \hat{\Psi}(x) \rangle$ and angular momentum distribution(lower panel) 
    for barrier height $\lambda = 
    1000$. From left to right $\Omega = 0, 0.5, 1.0$, corresponding to red crosses in 
    Fig.~\ref{fig:CurrentXRotation}, and as used before $\gamma = 10$ for 11 particles.}
    \label{fig:OBprobabilities}
\end{figure}

\noindent where $\tau = 2 \pi R / v$ is the period of barrier rotation. This quantify the mean 
fraction of particles that go through the  counter direction of the barrier in its 
period, that is from $\pi R$ to $-\pi R$ indicated by the limits of integration taken. Therefore, 
if $\langle \rho_s \rangle (v)$ takes the value 1, means  a perfect superfluid since all the 
particles are flowing with velocity $-v$ in the rotating frame, that is, they remain at rest for a 
observer in the laboratory frame. Relations with other observables can be established, for 
instance, using the average momentum per particle
{\small
    \begin{equation}
        \langle \rho_s \rangle(v) = \left(1 - \frac{\langle \hat{p} \rangle}{mv} \right), \quad
        \hat{p} = - \frac{i \hbar}{N} \int_{-\pi R}^{\pi R} \!\!\!\!\!\!\!\! \mathrm{d}x \ 
        \Psi^{\dagger}(x) 
        \frac{\partial}{\partial x} \Psi(x) ,
        \label{eq:currentMomentum}
    \end{equation}
}

\noindent and a relation with the energy, by taking the derivative with respect to the barrier velocity
{\small
    \begin{equation}
        \langle \rho_s \rangle (v) =  \frac{1}{N m v} \frac{\partial E}{\partial v}, \quad
        E = \langle \mathcal{H} \rangle .
        \label{eq:currentEnergy}
    \end{equation}
}

The equation above can also be identified by the ratio between the moment of inertia of the 
atoms and the moment of inertia of a rigid body. Using $v = \omega R$, yields
{\small
\begin{equation}
    \langle \rho_s \rangle(\omega) = \frac{1}{N m R^2} \left( \frac{1}{\omega}
    \frac{\partial E}{\partial \omega} \right) = \frac{I(\omega)}{I_{\mathrm{cl}}}.
    \label{eq:currentInertia}
\end{equation}
}

Superfluidity fraction at rest (or simply superfluidity fraction), denoted here by $\langle \rho_s
\rangle_0$ can be defined by taking the limit of $v \rightarrow 0$ in any of the forms 
\eqref{eq:currentJ},\eqref{eq:currentMomentum},\eqref{eq:currentEnergy} or 
\eqref{eq:currentInertia} and was studied in this way in previous 
works \cite{Hadzibabic2010,RevModPhys.Legget1999,Leggett1970,Leggett1998}.
With the dimensionless system of units and parameters introduced in the end of section \ref{secII}, 
we have a suitable expression for numerical calculations
{\small
\begin{equation}
    \langle \rho_s \rangle (\Omega) = \left( \frac{1}{2 \pi^2 N \Omega} \frac{\partial 
    E}{\partial \Omega} \right) , \ \langle \rho_s \rangle_0 = \lim_{\Omega \rightarrow 0}
    \langle \rho_s \rangle (\Omega) .
    \label{eq:rho_dimensionless}
\end{equation}
}

Here we use the MCTDHB to find the ground state through imaginary time propagation for 
several parameters, and we first study the effect of rotation. Fig.~\ref{fig:CurrentXRotation} 
illustrate the behavior of the energy in panel (A) and the current fraction in panel (B) as 
function of dimensionless barrier frequency $\Omega$ for two different barrier heights, where the 
specific form used in Eq. \eqref{eq:labframe} was
{\small
    \begin{equation}
        U(x) =  \left\{
        \begin{array}{lcl}
        \displaystyle
        (\hbar \zeta \lambda) \cos^2{\left( \frac{x}{2 R \sigma} \right)} & \mathrm{if} \ &
        | x | \leq \pi R \sigma \\
        0 & \mathrm{if} \ & \ \pi R \geq | x | > \pi R \sigma
        \end{array}
        \right. ,
    \label{eq:barrierform}
    \end{equation}
}

\begin{figure*}[tbp]
    \centering
    \includegraphics[scale=0.92]{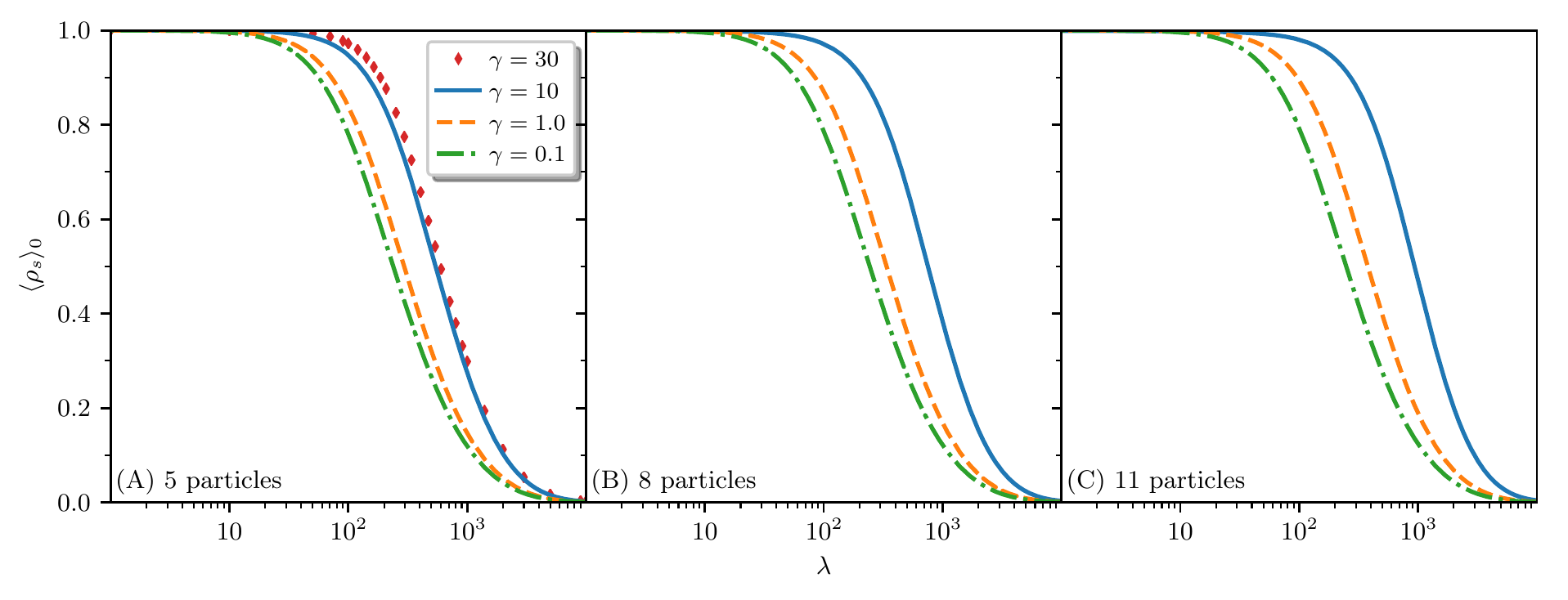}
    \caption{(Color online) Decrease of superfluidity fraction for different number of particles and 
interaction 
strength($\gamma$) due to increasing of the barrier height($\lambda$). All the cases share the 
common feature to be a perfect superfluid as the barrier becomes vanishing small, soon or later 
depending on the number of particles and interaction strength. For very high barriers
all particles are dragged together, imposing a rigid body rotation to the system.}
    \label{fig:flow_barrier}
\end{figure*}

\noindent where $\lambda$ denotes the barrier height in dimensionless units and the width of the 
barrier was taken fixed $\sigma = 0.1$~.

The energy of the ground state in Fig.~\ref{fig:CurrentXRotation}(A) has a period 1 with 
respect to dimensionless rotation frequency for both cases of weak and strong barriers, while the 
difference relies on the maximum that occurs at $\Omega_j = j$, that is peaked or smooth. The 
current fraction shows a periodic behavior with a damped amplitude as function of $\Omega$ in Fig.~\ref{fig:CurrentXRotation}(B), due to the periodicity of energy, where according to Eq. 
\eqref{eq:rho_dimensionless}, the amplitude is damped by a factor of $1 / \Omega$. In the regions 
where $\langle \rho_s \rangle (\Omega) < 1$ the average momentum must increase together with the 
barrier velocity by Eq. \eqref{eq:currentMomentum}. Indeed, that is what occurs in lower panel of 
Fig.~\ref{fig:OBprobabilities} that shows the angular momentum distribution fo some values of 
$\Omega$. Moreover, there is a critical dependence of the superfluid fraction on the barrier height, 
where Fig.~\ref{fig:CurrentXRotation} shows that, as $\Omega$ goes to zero, $\langle \rho_s \rangle 
(\Omega)$ becomes as smaller as higher is the barrier. This fact will be explored in the following.

\section{Decrease of superfluidity fraction due to increase of the barrier height}
\label{secIV}

Numerical calculations of superfluidity fraction was carried out here using Eq. 
\eqref{eq:currentEnergy}, finding the ground state by imaginary time propagation for $\Omega = 
0$ and $\Omega = 0.02$, to approximate the derivative in $\Omega = 0.01$ and so get $\langle 
\rho_s \rangle(0.01)$. As showed by Fig.~\ref{fig:CurrentXRotation} the slope of current 
fraction goes to zero as $\Omega \rightarrow 0$, and therefore we use the value at $\Omega = 0.01$ 
as the proper superfluidity fraction, assuming the difference of $\langle \rho_s \rangle_0 - 
\langle \rho_s \rangle(0.01)$ to be close to zero. To assure this method is valid, we 
compare with the result using Eq. \eqref{eq:currentMomentum} at $\Omega = 0.02$, to check if 
there is no appreciable(less than 1\%) variation on the estimation of superfluidity fraction using 
a constant extrapolation of $\langle \rho_s \rangle(0.01)$.

In Fig.~\ref{fig:flow_barrier}, we show the decrease of superfluid fraction for an increase 
in the barrier height for the form in Eq. \eqref{eq:barrierform}, using different number of 
particles and interaction strength. Here the tunneling of particles through the barrier is as harder 
as higher is the barrier, thereby the system acquires momentum easily for stronger barriers because 
it drags almost every particle with it. This easy momentum gain for very strong barriers is  
responsible for the loss of superfluidity fraction $\langle \rho_s \rangle_0$. The superfluidity fraction decreases more rapidly for 
fewer particles and lower interaction strength, however, the number of particles and strenght of 
interactions have a small impact in the form of the curves of $\langle \rho_s \rangle_0$ as a 
function of  $\lambda$.

\begin{figure}[tbp]
    \centering
    \includegraphics[scale=0.96]{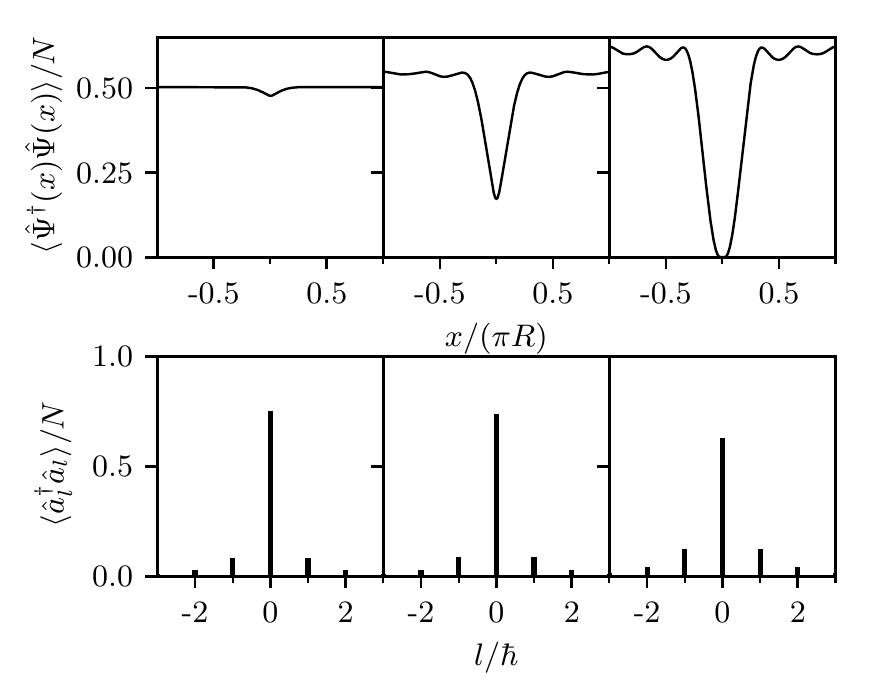}
    \caption{Probability distribution of position(upper panel) and angular momentum(lower panel) 
    for 5 particles, $\gamma = 30$ and different barrier heights $\lambda = 10, 200, 10000$ in the 
    left, center and right column respectively, in the absence of rotation $\Omega = 0$. The 
    density distribution vanishes for $\lambda > 10^3$ at the peak of the barrier in $x = 0$, 
    despite there is just a slight increase on the width of the angular momentum distribution.}
    \label{fig:OBprobabilities_barrier}
\end{figure}

As can be seen in upper panel of Fig.~\ref{fig:OBprobabilities_barrier}, the barrier height 
$\lambda$ influence mostly the density at its peak, while the effect over the momentum distribution 
is a slight increase in its variance, but preserving $\langle \hat{L} \rangle = 0$, as can be checked 
by the lower panel. This effect can also be seen in Fig.~\ref{fig:CurrentXRotation}, from which for 
$\lambda = 10$ the barrier has a critical value $\Omega = 0.5$ to start to move the particles, 
whereas for $\lambda = 1000$ it drags more easily the atoms since there $\langle \rho_s \rangle_0 
\approx 0.46$.

\begin{table}[hbt]

\begin{tabular}{c || @{\hskip 0.2cm} r |  @{\hskip 0.2cm} r |  @{\hskip 0.2cm} r |}
\backslashbox{$N$}{$\gamma$} & 1 & 10 & 30 \\ \hline \hline
11 & 0.9936 / 0.9920 & 0.92 / 0.89 & - \\ \hline
8  & 0.9946 / 0.9935 & 0.92 / 0.88 & - \\ \hline
5  & 0.9962 / 0.9956 & 0.91 / 0.88 & 0.75 / 0.70
\end{tabular}

\caption{maximum/minimum condensation fraction numbers given by the highest eingenvalue of 
$n^{(1)}(x,x')$, over the set of values of $\lambda$ in Fig.~\ref{fig:flow_barrier}. The maximum 
and minimum values for each case have little influence from the barrier height whereas the 
superfluid fraction maximum and minimum values goes from 1 to 0 respectively. For $\gamma = 30$ 
we were able to perform the calculations only for 5 particles due to our code
limitations.}

\label{tab:condensation_fraction}

\end{table}

\begin{figure}[tbp]
    \centering
    \includegraphics[scale=1.05]{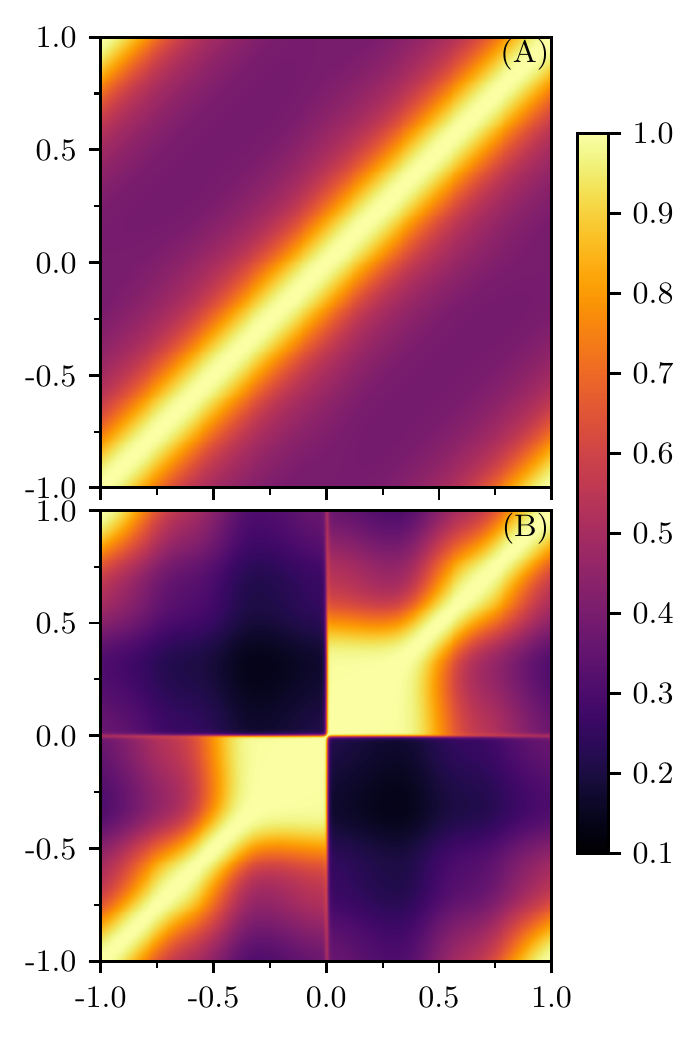}
    \caption{(Color online) $| g^{(1)}(x,x') |^2$ mapped to colors in the ring, for 5 particles and $\gamma = 
    30$, where the horizontal and vertical axes represent $x$ and $x'$ values in units of $\pi R$, 
    respectively. Values of barrier height used are $\lambda = 10$ in (A) and $\lambda = 10^4$ in 
    (B) but sharing the same color scale.}
    \label{fig:OBcorrelation}
\end{figure}

It is worth noting that this loss of superfluidity fraction due to increase in the barrier 
height is not related with the condensation fraction. As can be inferred from table
\ref{tab:condensation_fraction}, the condensation fraction depends mostly on the interaction strength and is minimally 
affected by the barrier height, particularly for small values of $\gamma$.

Being able to investigate many-body quantities that are unreachable using the GP equation, we 
further investigate how the tunneling amplitude is affected by the barrier height, that is the 
transition amplitude for the system to move a particle from $x$ to $x'$. This can be achieved by $| 
\langle \hat{\Psi}^{\dagger}(x') \hat{\Psi}(x) \rangle |^2$ weighted by the probabilities to find 
the particles in respective positions given by $n^{(1)}(x)$ and $n^{(1)}(x')$. This is directly 
related to the first order normalized correlation function defined by \cite{Glauber,PhysRevA.78.023615}
\begin{equation}
    g^{(1)}(x,x') =
    \frac{ \langle \hat{\Psi}^{\dagger}(x') \hat{\Psi}(x) \rangle }{\sqrt{n^{(1)}(x) 
    n^{(1)}(x')}} .
    \label{eq:correlation1}
\end{equation}

The values of $g^{(1)}$ shall be drastically affected by the barrier and must have an abrupt 
variation as $x x' > 0$ changes to $x x' < 0$, since the tunneling must be much harder 
if the shorter distance between two points has the barrier between them. Reminding that the system 
is periodic, this discussion applies just at the vicinity of either $x$ or $x'$ being zero, because 
if $x x' = -1$ they are actually the same point in the ring.

The effect of barrier height mentioned above is in agreement with the images in Fig.~\ref{fig:OBcorrelation} that maps $|g^{(1)}(x,x')|^2$ values to colors. In panel (A), in the 
presence of weak barrier, it depends only on $|x - x'|$, that means the tunneling is smaller as 
higher is the distance between the two points, while in panel (B) this symmetry is lost, with an 
abrupt variation near at the barrier peak, $x$ or $x'$ approximately zero. Therefore, high barriers 
split the image in four square blocks, with the darker regions (small normalized tunneling 
probabilities) located on $x x' < 0$. This is consistent with previous studies in Ref.~\cite{PhysRevA.78.013604}, despite the different boundary conditions and interaction regimes.

We further stress the relevance of applying a method that allows us to compute many-body 
quantities, since for example $| g^{(1)}(x,x') |^2$ would be identical to $1$ for all $x$ and $x'$ 
in case one uses the GP equation, that correspond to considering just one eigenstate of 
$n^{(1)}(x,x')$.

\section{Conclusions}
\label{secV}

In this paper, from a general derivation of the number of atoms current in a ring, we studied 
the persistent flow under the rotation of a barrier, and explained its behavior under different 
conditions using MCTDHB, which allowed us to explore strong interaction regimes with few particles. 
This method enable us to check convergence for the observables presented here, enlarging the basis 
of the spanned space by using more single-particle states determined by action minimization and,
therefore, achieve a beyond mean field theory and access to new observables as
the one-body correlation function.

Here we reported the periodicity of ground state energy under the MCTDHB and the effect 
of a barrier in the rotating frame, which changes the landscape of energy with respect to rotation 
velocity, producing narrow or smooth peaks periodically at dimensionless rotation frequency 
$\Omega_j = (j + 1/2)$ for weak or strong barriers respectively. The barrier also affects the 
particle number current fraction that either remains a perfect superfluid($\langle \rho_s \rangle 
(\Omega) = 1$) in the region $| \Omega | < 0.5$ for weak barriers, or has a fraction of the 
particles dragged by the system even for infinitesimal rotations, that is $\lim_{\Omega \rightarrow 
0}\langle \rho_s \rangle (\Omega) < 1$.

The superfluidity fraction decrease due to increase of the barrier height showed to be unrelated 
to condensation fraction since the value slightly changed under a wide range of values for the 
height of the barrier. However, the one body correlation function, that introduces a tunneling 
amplitude between two points weighted by the probability to find particles in the respective points 
is a key quantity to understand this observation. If the particles can pass through the obstacle 
without gaining momentum, in other words, tunnel through the barrier, this flow behaves as a perfect 
superfluid, that is,  it stays at rest as the barrier starts to move. Indeed, this was 
quantitatively predicted by the one-body correlation function.

\section*{Acknowledgements}
The authors thank the Brazilian agencies Funda\c{c}\~ao de Amparo \`a Pesquisa do Estado de S\~ao Paulo
(FAPESP) and Conselho Nacional de Desenvolvimento Cient\'ifico e Tecnol\'ogico (CNPq). We gratefully 
thanks to A. F. R. T. Piza, E. J. V. Passos and R. K. Kumar for the elucidating discussions. 
We are also grateful for conversations with A. U. J. Lode and M. C. Tsatsos about the 
implementation of codes to numerically solve the MCTDHB equations.

    \bibliographystyle{apsrev4-1}
    \bibliography{ref}
    
\end{document}